# Hybrid MPI-OpenMP Paradigm on SMP clusters: MPEG-2 Encoder and n-body Simulation

Truong Vinh Truong Duy, Katsuhiro Yamazaki, Kosai Ikegami, and Shigeru Oyanagi

*Abstract*— **Clusters of SMP nodes provide support for a wide diversity of parallel programming paradigms. Combining both shared memory and message passing parallelization within the same application, the hybrid MPI-OpenMP paradigm is an emerging trend for parallel programming to fully exploit distributed shared-memory architecture. In this paper, we improve the performance of MPEG-2 encoder and n-body simulation by employing the hybrid MPI-OpenMP programming paradigm on SMP clusters. The hierarchical image data structure of MPEG bit-stream is eminently suitable for the hybrid model to achieve multiple levels of parallelism: MPI for parallelism at the group of pictures level across SMP nodes and OpenMP for parallelism within pictures at the slice level within each SMP node. Similarly, the work load of the force calculation which accounts for upwards of 90% of the cycles in typical computations in n-body simulation is shared among OpenMP threads after ORB domain decomposition among MPI processes. Besides, loop scheduling of OpenMP threads is adopted with appropriate chunk size to provide better load balance of work, leading to enhanced performance. With n-body simulation, experimental results demonstrate that the hybrid MPI-OpenMP program outperforms the corresponding pure MPI program by average factors of 1.52 on 4-way cluster and 1.21 on 2-way cluster. Likewise, the hybrid model offers a significant performance improvement of 18% compared to MPI model for the MPEG-2 encoder.**

## I. INTRODUCTION

LARGE scale highly parallel systems based on cluster of SMP architecture are today's dominant computing platforms which enable many different parallel programming paradigms. Optimal paradigms enable application developers to use the hardware architecture in a most efficient way, i.e., without any overhead induced by the programming paradigm. On distributed memory systems, MPI [9] is widely used for writing message passing programs across the nodes of a cluster while OpenMP [10] is a popular API for parallel programming on shared memory architecture. As a result, a combination of shared memory and message passing parallelization paradigms within the same application, hybrid programming, is expected to provide a more efficient parallelization strategy for clusters of SMP nodes. The hybrid MPI-OpenMP approach supports multiple levels of parallelism on an SMP cluster where MPI is used to handle parallelism across nodes and OpenMP is employed to exploit parallelism within a node.

There have been many efforts for porting message passing applications to hybrid applications, leaving both opportunities and challenges of getting higher performance with this model. The implementation, development and performance of hybrid program applications are discussed in [2]. The results demonstrate that this style of programming is not always be the most effective mechanism but can obtain significant benefits from some situations. Similarly, the results from comparing MPI with MPI-OpenMP for the NAS benchmarks [3] are clearly application-dependent. The hybrid approach becomes better when processors make the communication performance considerable and the level of parallelization is sufficient. Bush et al. [6] prove that hybrid MPI-OpenMP codes can give significant performance on kernel algorithms such as Cannon's matrix multiply although it requires a substantial amount of work involved to achieve this. The performance of hybrid message-passing and shared-memory parallelism for discrete element modeling is presented in [4]. However, the authors conclude their current OpenMP implementation is not yet efficient enough for hybrid parallelism to outperform pure message-passing on an SMP cluster although OpenMP is more effective than MPI on a single SMP node.

The aim of this paper is to improve the performance of MPEG-2 encoder and n-body simulation by employing the hybrid MPI-OpenMP programming paradigm on SMP clusters. The performance of MPEG-2 encoding application is becoming more important as the demand for multimedia applications increases. Moreover, the hierarchy of layers in an MPEG-2 bit-stream is eminently suitable for applying the hybrid paradigm to exploit two levels of parallelism: parallelism at the group of pictures level and parallelism within pictures at the slice level. Meanwhile, the n-body is a classical one, and appears in many areas of science and engineering, including astrophysics, molecular dynamics, and graphics. In the simulation of n-body, the specific routines for calculating the force of the bodies which account for almost 90% of total computing time are very appropriate for parallelizing with OpenMP work-sharing directives. In addition, we removed unnecessary MPI intra-node communication and employed loop scheduling of OpenMP

Graduate School of Science and Engineering, Ritsumeikan University, Shiga, Japan (e-mail: duy@hpc.cs.ritsumei.ac.jp).

threads with appropriate chunk size for better load balance in the hybrid model. The rest of the paper is structured as follows. Section 2 presents the hybrid MPI-OpenMP parallel programming paradigm. Parallelization and implementation of the MPEG-2 encoding application are detailed in section 3 while the n-body simulation is described in section 4. Section 5 analyzes the experimental results. Finally, we conclude our study in section 6.

## II. Hybrid MPI-OpenMP programming paradigm

Often, hybrid MPI-OpenMP programming refers to a programming style in which communication between nodes is handled by MPI processes and each MPI process has several OpenMP threads running inside to occupy the CPUs of an SMP node. The number of OpenMP threads is equal to the number of CPUs in one SMP node and there are as many MPI processes as nodes in a cluster. However, this style, called process-to-process communication method, is only one in two main different hybrid programming styles characterized by whether OpenMP threads take part in communication between nodes or not. In process-to-process communication, MPI routines are invoked outside of OpenMP parallel regions, thus there is only MPI communication between nodes. On the other hand, in thread-to-thread communication, some MPI routines are placed inside of OpenMP parallel regions, leading to OpenMP threads' involvement in inter-node communication.

Each style has different merits and demerits [1], and appropriate for different classes of applications. For implementation of our applications, MPEG-2 encoder and n-body simulation, the process-to-process communication model is more efficient. An MPEG-2 bit-stream can be easily broken into independent groups of pictures and MPI processes are used to carry out these groups. Each picture in turn can be processed in parallel with multiple OpenMP threads running inside MPI process. For the n-body simulation, parallelizing the specific time-consuming routines using lighter-weight OpenMP threads without having to communicate with each other is more effective.

Furthermore, this method outweighs other methods in terms of communication overhead. It requires only inter-node communication between nodes since intra-node communication is substituted by direct access to the shared memory. Meanwhile, for example with pure MPI, additional intra-node communication is necessary within each node between MPI processes as illustrated in Figure 1.

On the other hand, in OpenMP loop parallelization of hybrid MPI-OpenMP programs, there is no guarantee that just because a loop has been correctly parallelized, its performance will improve. In fact, in some circumstances parallelizing the wrong loop can slow the program down. Even when the choice of loop is reasonable, some performance tuning may be necessary to make the loop run acceptably fast. Hence, among several mechanisms for controlling this factor provided by OpenMP, we employed loop scheduling to ensure sufficient work with better load balance for OpenMP threads.

## III. MPEG-2 encoding application

### A. MPEG Overview

MPEG is an encoding and compression system for digital multimedia content defined by the Motion Pictures Expert Group [15]. The MPEG-2 video compression algorithm achieves very high rates of compression by removing both the temporal redundancy and spatial redundancy present in motion video. An important aspect of MPEG which can take advantage of the hybrid parallel programming is its layered structure. The hierarchy of layers in an MPEG bit-stream is arranged in the following order: Sequence, Group of Pictures (GOP), Picture, Slice, Macro-block, and Block (Figure 2).

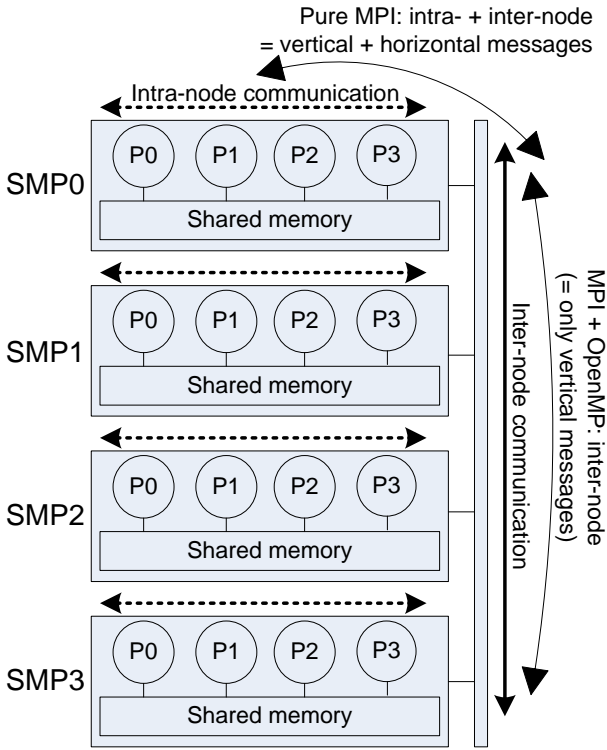

Fig. 1. Communication pattern in MPI and hybrid programming models.

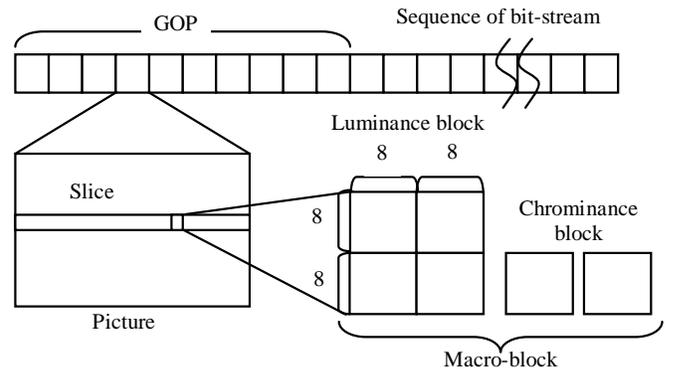

Fig. 2. The hierarchy of layers in an MPEG bit-stream.

The highest level in the layering is the sequence level. A sequence is made up of groups of pictures (GOPs). Each GOP is a grouping of a number of adjacent pictures. Pictures (frames) are further subdivided into slices, each of which defines a fragment of a row in the picture. Slices comprise a series of macro-blocks, which are 16x16 pixel groups containing the luminance and chrominance data for those pixels in the decoded picture. Macro-blocks are divided into blocks (6 to 12 depending upon format). A block is an 8x8 pixel group that describes the luminance or chrominance for that group of pixels. Blocks are the basic unit of data at which the decoder processes the encoded video stream.

The basic operation of an MPEG-2 encoder is described briefly in Figure 3. Two key techniques are intra-frame DCT coding for removing spatial redundancy and inter-frame prediction (MCP) for exploiting temporal redundancy by attempting to predict the frame to be coded from a previous 'reference' frame. The coder subtracts the motion-compensated prediction from the source picture to form a 'prediction error' picture that represents the difference between the predicted macro-block and the actual macro-block being encoded. The prediction error is transformed with the DCT to produce blocks of DCT coefficients. The DCT coefficients are then quantized at Q to reduce the number of bits needed to represent each coefficient. The quantized DCT coefficients are Huffman coded in combination with run-level coding and zig-zag scanning using a VLC which further reduces the average number of bits per coefficient. This is combined with motion vector data and other side information, and formed into a coded bit-stream out.

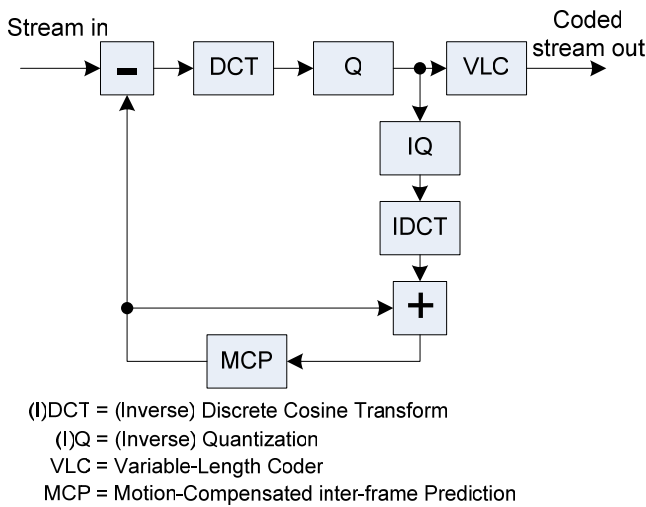

Fig. 3. Basic operation of MPEG-2 encoder.

The quantized DCT coefficients after Q also go to an internal loop to be inverse quantized (IQ) and inverse DCT transformed (IDCT). The predicted macro-block read out of the reference picture memory is added back to the residual on a pixel by pixel basis and stored back into memory to serve as MCP's reference for predicting subsequent pictures. Another object is to have the data in the reference picture memory of the encoder match the data in that of the decoder.

## B. Exploiting Two Levels of Parallelism

The hierarchical image data structure of MPEG bit-stream is eminently suitable for using the hybrid paradigm to apply parallelism beyond a single level. The possible choices for a task in MPEG are: Sequence, Group of Pictures (GOP), Picture, Slice, Macro-block, and Block. For the first level, parallelizing across sequences may lead to tasks which are too large and create load imbalance. Consequently, a more reasonable choice is to parallelize across GOPs. Since GOPs are relatively independent, there is essentially no inherent communication in the parallel algorithm. For the second level, assigning adjacent pictures to different tasks leads to many serializing dependencies, and associated synchronization among the tasks because one picture depends on other nearby pictures. Meanwhile, dividing tasks at macro-block or block level results in tiny tasks. Therefore, parallelism within a picture at slice level is the most appropriate approach.

*1) Parallelism at GOP level:* At this level, the incoming stream is decomposed into GOPs which are assigned to MPI process. We implemented two styles of partitioning: block and cyclic. In the former, GOPs are divided into as many relatively equal blocks as MPI processes, and each carries out one block. In the latter, GOPs are distributed to MPI processes in a fashion similar to round-robin: the first process gets the first GOP, the second process gets the second GOP, and so on, until no more GOPs remain as shown in Figure 4.

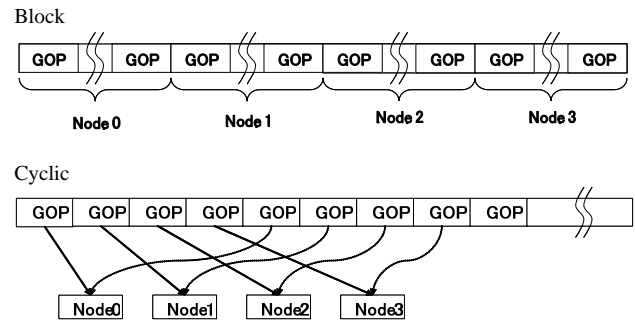

Fig. 4. Block and cyclic partitioning of GOPs

*2) Parallelism at slice level:* Parallelism at slice level is achieved by using multiple OpenMP threads running inside MPI to process slices within a picture in parallel. Synchronization among threads necessary at the end of every picture is trivial because the pictures have been loaded into the shared-memory.

Our parallel implementations of MPEG-2 encoder are based on the corresponding sequential program provided by MPEG group [16]. The pseudo code of the hybrid program in which OpenMP directives are inserted in the MPI implementation for the main loop processing each picture is as follows:

```
{
MPI_Initialize();
…
MPI_Scatter(MPI_processes, GOPs);
…
For each picture in GOP{
```

```
#pragma omp parallel for private (n)
//The work load here is divided among OpenMP
threads
For slice#0 to slice#n {Process_slice;}
}
…
MPI_Gather(MPI_processes, output);
…
MPI_Finalize();
}
```

## IV. THE N-BODY SIMULATION

The n-body problem involves advancing the trajectories of n bodies according to their time evolving mutual gravitation field. The Barnes-Hut method [16], or tree code algorithm, which has advantage of scaling only as O(NlogN) in computational cost, is widely used to solve this problem today.

### A. Tree Code Algorithm

The essence of the tree code is the recognition that a distant group of bodies can be well-approximated by a single body, located at the center of mass with a mass equal to total mass of the group. It represents the distribution of the bodies in quad-tree for 2D space or oct-tree for 3D space. The tree is implemented by recursively dividing the 2D space into 4 subspaces, or 8 subspaces in 3D space, until the number of bodies in each subspace is below a certain threshold. Figure 5 demonstrates the distribution of bodies in 2D space and the corresponding quad-tree.

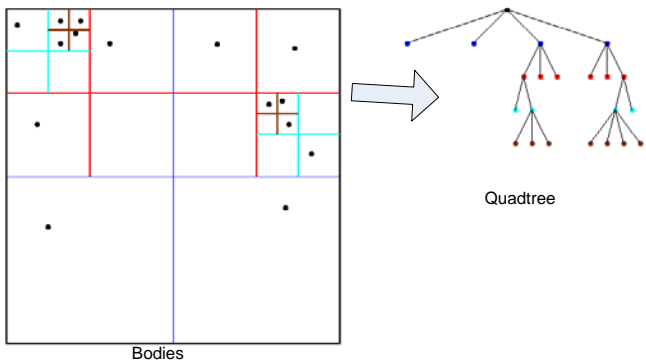

Fig. 5. Bodies in 2D space and the quad-tree.

After the tree construction phase, the force on a body in the system can be evaluated by traversing down the tree from root. At each level, a cell is added to an interaction list if the cell is distant enough for a force evaluation. Otherwise, the traversal continues recursively with the children. The accumulated list of interacting cells and bodies is then looped through to calculate the force on the given body. Finally, each body updates its position and velocity based on the computed forces.

### B. Parallelization of Tree Code

The sequential algorithm works well but there is a serious problem in parallelization of tree code. The tree is very unbalanced since the bodies are not uniformly distributed in their bounding box. Hence, it is important to divide space into domains with equal work-loads to avoid load imbalance. We adopted the Orthogonal Recursive Bisection (ORB) domain decomposition [18] to divide the space into as many non-overlapping subspaces as processors, each of which contains an approximately equal number of bodies, and assign each subspace to a processor. Figure 6 shows the ORB domain decomposition in 2D space on 16 processors.

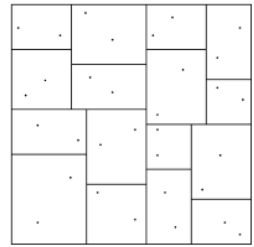

Fig. 6. ORB domain decomposition in 2D space on 16 processors.

Furthermore, in order for a processor to run tree code for computing forces of its own bodies, each processor builds a local tree for its set of bodies which is later extended into a Locally Essential Tree (LET). LET contains all the nodes of the global tree that are essential for the bodies contained within that processor. Each processor computes the destination processors for which the node might be essential; this involves the intersection of the annular region of influence of the node, called "influence ring", with the ORB map as illustrated in Figure 7. Those bodies that are not in the influence ring are either too close to node *u* to apply center-of-mass approximation, or far away enough to use u's parent's information, therefore *u* will be essential to only particles within its influence ring. As a result, each processor first collects all the information deemed essential to other nodes, and then exchanges data directly with the appropriate destinations. Once all processors have received and inserted the data received into the local tree, each processor has its own LET. Then, every processor can proceed exactly as in the sequential case.

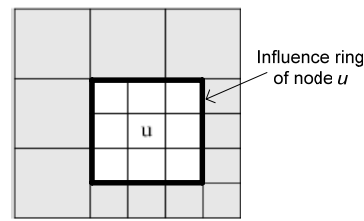

Fig. 7. The influence ring of a node *u* in the ORB map.

The parallel programs implementing the tree code have been developed from the sequential tree code provided by Joshua E. Barnes [17] in which the force calculation accounts for upwards of 90% of the cycles in typical computations. By using ORB domain decomposition to distribute the bodies in a balanced way among MPI processes and OpenMP threads for sharing work load of the force calculation in each process, the hybrid program is expected to speed up the performance. The following section briefly presents pseudo code of the hybrid program.

```
{
MPI_Initialize();
…
ORB_domain_decomposition(MPI_processes,
bodies);
Constructs_the_local_tree_code(my_bodies);
Build_the_LET(MPI_processes);
…
#pragma omp parallel for private (n) schedule
(type)
//The work load here is divided among OpenMP
threads
For body#0 to body#n in interaction list
{
Calculate_forces();
}
…
Move_bodies(my_bodies);
…
MPI_Finalize();
}
```

## V. Experimental results

In this section, we describe timing runs of the codes taken on our SMP clusters focusing on performance with execution time as the metric for measurements. First we give a description of the compute platforms. Then the performance evaluation results of MPEG-2 encoder and n-body simulation are discussed.

### A. Compute Platforms

System specifications of two SMP clusters used for evaluating the MPI and hybrid MPI-OpenMP programs are detailed in Table I. The codes are compiled using Intel C Compiler and MPI library of MPICH implementation. On these systems, we repeated the experiments 5 times and observed small performance variations with standard deviation for 5 measures less than 5%. The average of these 5 measures is presented.

TABLE I
System specifications

| Name | SMP Node | # of Nodes | # of CPUs | Network |
|---|---|---|---|---|
| Diplo | Quad Xeon 3GHz | 4 | 16 | Gigabit Ethernet |
| Atlantis | Dual Xeon 2.8GHz | 16 | 32 | Gigabit Ethernet |

### B. The MPEG-2 Encoder

The sample stream used for evaluating the MPEG-2 encoding application is a group of animated pictures of natural scenery, a forest, where the movement is little. It consists of 1920 frames with a length of 64 seconds in 3 different levels for the Main Profile: the High1440 Level (MP@HL), Main Level (MP@ML), and Low Level (MP@LL). Table II shows characteristics of each level varying resolution, number of frames per second, and bit rate.

TABLE II
Characteristics of different levels for MPEG-2 MP

| Level | Resolution | Frames per second | Max bit rate |
|---|---|---|---|
| Low (MP@LL) | 352 x 288 | 30 | 4Mb/s |
| Main (MP@ML) | 720 x 576 | 30 | 15Mb/s |
| High 1440 (MP@HL) | 1440 x 1088 | 60 | 60Mb/s |

The performance of MPEG-2 encoder for those varying MPEG-2 levels run on the 2-way Atlantis cluster is displayed in Table III and Figure 8. Clearly, the hybrid programs are better than the pure MPI programs in most cases whatever processors and data sets are used. The benefits of multiple levels of parallelism offer a significant performance improvement of 18% for hybrid programs. We also noted that when the test stream with little movement is dealt with, for example images of a forest, it is possible to further shorten execution time with hybrid codes in comparison with MPI codes. Because the color is almost similar in scenery pictures of such little movement streams, the compression ratio is high. In contrast, this ratio is lower when processing complicated pictures of quick movements such as the sea waves. Namely, since the picture compression is done in the shared-memory with parallel OpenMP threads in the hybrid implementations, it is thought that the hybrid execution is becoming even more effective when the compressibility of still pictures is high.

TABLE III
Execution time of MPEG-2 encoder on 2-way atlantis

| | # of CPUs | | 2 | 4 | 8 | 16 | 32 |
|---|---|---|---|---|---|---|---|
| MP@HL (minutes) | Block | Hybrid | 100.2 | 51.6 | 25.7 | 13.3 | 7.4 |
| | | MPI | 116.9 | 59.2 | 29.9 | 15.1 | 8.5 |
| | Cyclic | Hybrid | 105.7 | 54.1 | 26.9 | 13.7 | 7.6 |
| | | MPI | 119 | 61.2 | 31.1 | 16 | 8.6 |
| MP@ML (minutes) | Block | Hybrid | 20.8 | 10.5 | 5.4 | 3.1 | 2.6 |
| | | MPI | 23.8 | 12 | 6.5 | 3.9 | 2.7 |
| | Cyclic | Hybrid | 21.9 | 11 | 5.6 | 3.1 | 2.5 |
| | | MPI | 25 | 12.7 | 6.6 | 4.1 | 2.6 |
| MP@LL (seconds) | Block | Hybrid | 291.4 | 150.5 | 80.1 | 62.8 | 70.9 |
| | | MPI | 349.7 | 177.8 | 93.5 | 65.2 | 68.2 |
| | Cyclic | Hybrid | 315.8 | 158.9 | 85.6 | 60.1 | 63.9 |
| | | MPI | 372 | 183.1 | 97.5 | 63.4 | 63.7 |

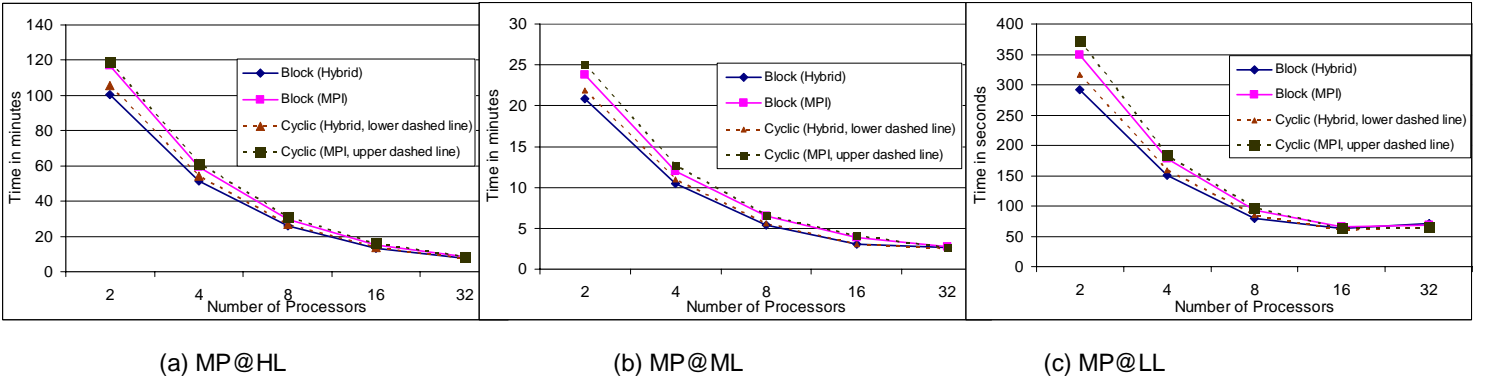

(a) MP@HL      (b) MP@ML      (c) MP@LL

Fig. 8. Execution time of MPEG-2 encoder with MP@HL, MP@ML, MP@LL on 2-way Atlantis.

In general, both pure MPI codes and hybrid codes scale well. However, in case of ML, speed improvement ratio decreases with 32 processors as against 16 processors (Figure 8b). With LL the largest speed improvement has been achieved with 16 CPUs (Figure 8c). In case of 32 processors, with data sizes of ML and LL equal to 120MB and 36MB respectively, the effect of hybrid implementations slides with the data size smaller than 120MB, and LL hybrid execution has become slower than MPI execution with 32MB. Therefore, it is necessary to modify divided grain size according to the number of CPUs to maintain the efficiency of the hybrid codes.

Concerning partitioning strategy, we remarked that the block partitioning outweighs cyclic partitioning with up to 8 CPUs in all data sizes and the programs employing block partitioning are on average 5% faster than cyclic programs. The cyclic partitioning begins to take effect in ML with 32 CPUs and in LL with 16 or more processors. Table IV presents execution time of the hybrid programs with different minimum units of cyclic partitioning compared to block partitioning using ML data size. With cyclic partitioning, we designated the minimum unit of division varying from 1 to 4 GOPs, and the result with 4 GOPs is best for 32 processors whereas the block partitioning exhibits the best performance with 16 CPUs. However, as the number of frames inside GOP is variable, from several dozens to 100 frames, examination of the optimal unit of division, the number of frames, and other related issues is important.

TABLE IV
EXECUTION TIME OF CYCLIC PARTITIONING (SECONDS)

| # OF CPUs | 2 | 4 | 8 | 16 | 32 |
|---|---|---|---|---|---|
| 1GOP | 1314.7 | 661.9 | 338.4 | 187.3 | 150.3 |
| 2GOP | 1366.2 | 689.8 | 361.0 | 208.2 | 149.1 |
| 4GOP | 1362.9 | 683.4 | 355.6 | 195.0 | 147.0 |
| Block | 1248.5 | 629.3 | 322.5 | 184.9 | 158.0 |

## C. The n-body Simulation

The n-body programs simulating the interactions among $10^5$ bodies in 10 timesteps have been run on the 4-way Diplo cluster and 2-way Atlantis cluster. Figure 9 and Table V display the execution time of the MPI and hybrid codes for a fixed number (1, 2, 4, 8, 16, and 32) of CPUs.

TABLE V
EXECUTION TIME OF $10^5$-BODY SIMULATION (IN MINUTES)

| | # OF CPUs | 1 | 2 | 4 | 8 | 16 | 32 |
|---|---|---|---|---|---|---|---|
| Diplo | Hybrid | – | 48.6 | 24.9 | 12.6 | 6.4 | – |
| | MPI | 134.5 | 71.6 | 37.8 | 19.7 | 10.4 | – |
| Atlantis | Hybrid | – | 54.3 | 27.1 | 14.8 | 7.6 | 4 |
| | MPI | 124.3 | 63 | 31.9 | 16.2 | 8.3 | 4.3 |

In both clusters, no matter how many processors are used, the hybrid implementation outperforms the pure MPI one by average factors of 1.52 on Diplo and 1.21 on Atlantis at all times. We also observed that the factor on Diplo is higher than on Atlantis, resulting from a greater number of OpenMP threads on Diplo in contrast to Atlantis because there are 4 OpenMP threads created on a 4-way compute node, and only 2 OpenMP threads on a 2-way compute node for each MPI process to best suit the system architecture. As a result, it is expected that this factor will be even higher in 8 or 16-way clusters although we have not had the opportunity to test the codes in such systems.

Besides, another important advantage of the hybrid model compared to pure MPI model is that it lowers the number of sub-domains in ORB domain decomposition. For instance, we need to create only 4 sub-domains for the hybrid program while 16 sub-domains are necessary for the MPI program on 4-way Diplo cluster. As the number of sub-domains increases, the shapes of domains have a larger range of aspect ratios forcing tree walks to proceed to deeper levels. The complexity involved in determining locally essential data also rises with the number of sub-domains. We found that the number of cell interactions grows with the number of sub-domains because of these effects. Thus, the hybrid model helps reduce this cell interaction overhead.

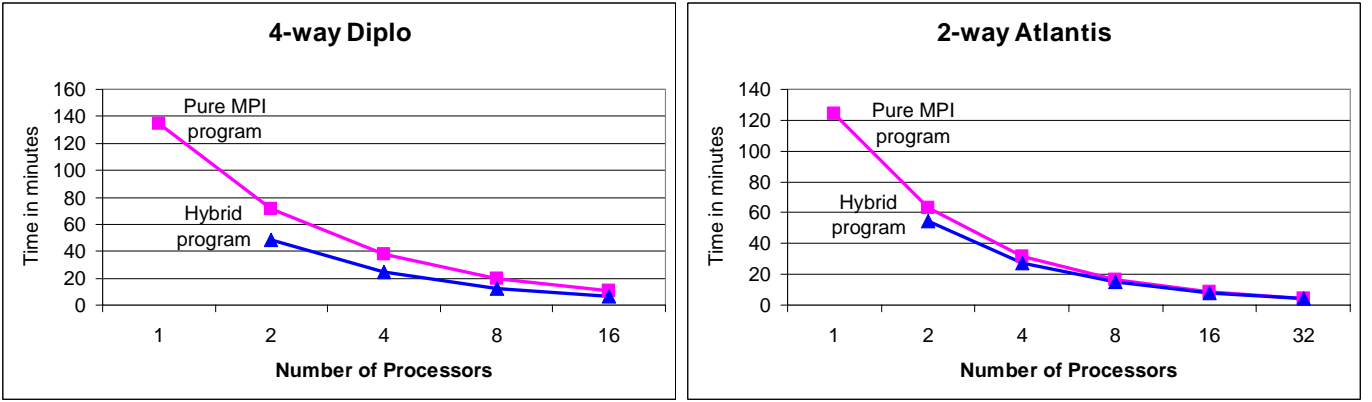

Fig. 9. Execution time of $10^5$-body simulation on 4-way Diplo and 2-way Atlantis clusters.

However, there is a main disadvantage that exists in both models. The memory requirements of the parallel tree code are quite large. The majority of memory is assigned to the body and cell arrays for the local subset of bodies with additional arrays for bodies and cells imported for the locally essential trees. On modern computing platforms with large amount of memory, it seems no longer a problem. We did not encounter any problem with memory on our systems. We also tested performance of the hybrid program with a combination of different scheduling methods and chunk sizes on 4-way cluster using 8 and 16 CPUs. The timing results obtained by running the hybrid code are listed in Table VI.

TABLE VI
EXECUTION TIME WITH DIFFERENT SCHEDULES AND CHUNK SIZES (SECONDS)

| | CHUNK SIZE | 1 | 100 | 500 | 1000 | 2000 | 5000 |
|---|---|---|---|---|---|---|---|
| 8 CPUs | Static | 756.7 | 755.7 | 754.8 | 757 | 752.3 | 906.8 |
| | Dynamic | 755.2 | 754.3 | 751.8 | 756.2 | 752.1 | 906.8 |
| | Guided | 754.1 | 758 | 765.5 | 756.3 | 767.7 | 908.8 |
| 16 CPUs | Static | 385.9 | 389.9 | 387.8 | 385.7 | 409.8 | 447.4 |
| | Dynamic | 385.2 | 388.2 | 386.3 | 384.1 | 408.1 | 445.3 |
| | Guided | 385.6 | 387.1 | 388.7 | 390.6 | 407.9 | 445.6 |

The tested schedules are:

-- Static: Loop iterations are divided into pieces of size *chunk* and then statically assigned to threads in a round-robin fashion.

-- Dynamic: Loop iterations are divided into pieces of size *chunk*, and dynamically scheduled among the threads; when a thread finishes one chunk, it is dynamically assigned another.

-- Guided: For a chunk size of 1, the size of each chunk is proportional to the number of unassigned iterations divided by the number of threads, decreasing to 1. For a chunk with value k greater than 1, the size of each chunk is determined in the same way with the restriction that the chunks do not contain fewer than k iterations (except for the last chunk to be assigned, which may have fewer than k iterations).

From Table VI, it is easy to recognize that schedule dynamic outweighs schedule static and guided in most cases, and the chunk size has an important impact on the performance. As mentioned earlier, the routine calculating the force of bodies accounts for the vast majority of the cycles in typical calculations in this simulation. Synchronization overhead incurred by dynamic scheduling is trivial beside this computation time. Consequently, schedule dynamic is always better than schedule static with all chunk sizes, and also provides better load balance than schedule guided in many cases even though the difference among schedule styles is quite small.

What made major difference here is the size of chunk used for scheduling. With chunk sizes equal to or less than 1000 (16 CPUs) and 2000 (8 CPUs), execution time has not changed much with all schedule types. The chunk sizes of 500 and 1000 are the best with dynamic scheduling in case of 8 and 16 CPUs respectively. The quality of load balance drops with increasing chunk size and running time grows steadily with chunk size greater than 1000 (16 CPUs) and 2000 (8 CPUs). As the bodies are not uniformly distributed in their bounding box, the force computation time varies enormously from one body to another. Therefore, a chunk size which is too large easily leads to load imbalance. This means that choosing a chunk size is a trade-off between the quality of load balancing and the synchronization and computation costs.

## VI. Conclusion

In this paper, we studied the performance and the programming efforts for two different applications, MPEG-2 encoder and n-body simulation, under two parallel programming paradigms: pure MPI and hybrid MPI-OpenMP. With the hybrid model, multiple levels of parallelism can be achieved. Parallelization is performed at both the GOP and slice levels in the MPEG-2 encoding application. Likewise, the work load of time-consuming routines for calculating forces of the bodies in n-body simulation is shared among OpenMP threads after ORB domain decomposition among MPI processes. In addition, loop scheduling of OpenMP threads is employed with appropriate chunk size for better load balance in the hybrid program, resulting in enhanced performance. Given these abilities, the hybrid MPI-OpenMP programs outperform the corresponding pure MPI programs in terms of execution time for both applications in most cases whatever processors and data sets are used. Thus, this paper gives a positive aspect of developing hybrid MPI-OpenMP parallel paradigms for real applications. With respect to the achieved results, we believe that for some certain classes of problems, the hybrid paradigm provides the most efficient mechanism to fully exploit clusters of SMP nodes.

More experiments are necessary to evaluate the MPEG-2 encoder on 4-way Diplo cluster with the best combination of chunk size and number of processors. Besides, the encoder is actually only one component of a video codec consisting of an encoder and a decoder, which respectively performs compression and decompression of video data. A full parallel version of video codec with both the encoder and decoder are parallelized is worth considering. For solving the n-body problem, a number of methods have also been introduced in addition to Barnes-Hut in which the Fast Multipole Method (FMM) algorithm [20] has been shown to be O(N). It is expected that superior performance can be achieved by adapting the parallel tree code using FMM algorithm. So far, the experiments have been done only on separate SMP clusters. Therefore, porting these programs to a multiple SMP cluster computing environment is a proper approach to maximize the use of resources and provide much higher throughput.

## References

[1] R. Rabenseifner, "Hybrid Parallel Programming: Performance Problems and Chances", Proc. of the 45th Cray User Group Conference, 2003.
[2] L.Smith, M.Bulk, "Development of Mixed Mode MPI/OpenMP Applications", WOMPAT 2000, 2000.
[3] F. Cappello, and D. Etiemble, "MPI versus MPI+OpenMP on IBM SP for the NAS Benchmarks", Proc. of Supercomputing, 2000.
[4] D. S. Henty, "Performance of Hybrid Message-Passing and Shared-Memory Parallelism for Discrete Element Modeling", Proc. of Supercomputing, 2000.
[5] L.A. Smith, and P. Kent, "Development and performance of a mixed OpenMP/MPI Quantum Monte Carlo code", Concurrency: Practice and Experience 12 (2000), 1121–1129.
[6] I.J. Bush, C.J. Noble, and R.J. Allan, "Mixed OpenMP and MPI for Parallel Fortran Applications", http://www.ukhec.ac.uk/publications/reports/ewomp paper.pdf..
[7] A. Kneer, "Industrial Hybrid OpenMP/MPI CFD application for Practical Use in Free-surface Flow Calculations", WOMPAT2000: Workshop on OpenMP Applications and Tools, 2000.
[8] Yun He, and C. HQ Ding, "MPI and OpenMP paradigms on cluster of SMP architectures: the vacancy tracking algorithm for multi-dimensional array transposition", Supercomputing 2002.
[9] MPI, MPI: A Message-Passing Interface standard. Message Passing Interface Forum, June 1995, http://www.mpiforum.org/.
[10] OpenMP, The OpenMP ARB, http://www.openmp.org/.
[11] K. Yamazaki, K. Ikegami, and S. Oyanagi, "Speed Improvement of MPEG-2 Encoding using Hybrid Parallel Programming", IPSJ and IEICE, FIT 2006, Information Technology Letters, LC-005, Vol.5, 2006.
[12] E. Iwata, and K. Olukotun, "Exploiting Coarse-Grain Parallelism in the MPEG-2 Algorithm", CSL-TR-. 98-771, 1998.
[13] S.M. Akramullah, I. Ahmad, and M. L. Liou, "A data-parallel approach for real-time MPEG-2 video encoding", Journal of Parallel and Distributed Computing, 30(2):129–146, 1995.
[14] A. Bilas, J. Fritts, and J. P. Singh, "Real-Time Parallel MPEG-2 Decoding in Software," TR-516-96, Princeton University, 1996.
[15] MPEG, MPEG group, http://www.mpeg.org/MPEG/MSSG/
[16] J. Barnes, and P. Hut, "A hiearchical o(nlogn) force calculation algorithm", Nature, Vol.324, pp.446–449, 1986.
[17] Treecode, treecode guide, http://ifa.hawaii.edu/~barnes/treecode/treeguide.html
[18] J. Dubinski, "A parallel tree code", New Astronomy 1 (1996) 133-147.
[19] P. Mioochi, "Simulation of the dynamics of globular clusters: an efficient parallelization of a tree-code", Proc. of the conference "Dynamics of Star Clusters and the Milky Way", 2000.
[20] J. Makino, "Yet another fast multipole method without multipoles --- pseudo-particle multipole method", Journal of Computational Physics, Vol.151, pp.910--920, 1999.